\documentclass[multicol,prl,twocolumn,aps]{revtex4}
\usepackage{graphicx}% Include figure files
\usepackage{bm}% bold math
\usepackage{color}
\usepackage{amssymb}
\usepackage{latexsym}
\usepackage{amsmath}

\begin{document}

\title{Superconductivity in a two-dimensional superconductor with Rashba and Dresselhaus spin-orbit couplings}

\author{Xu Yan}

\author{Qiang Gu}
\email{qgu@ustb.edu.cn} \affiliation{Department of Physics,
University of Science and Technology Beijing, Beijing 100083, China}

\date{\today}

\begin{abstract}
We present a general model with both Rashba and Dresselhaus
spin-orbit couplings to describe a two-dimensional
noncentrosymmetric superconductor. The combined effects of the two
spin-orbit couplings on superconductivity are investigated in the
framework of mean-field theory. We find that the Rashba and
Dresselhaus spin-orbit couplings result in similar effects on
superconductivity if they are present solely in the system. Mixing
of spin-singlet and triplet pairings in electron band is induced
under the assumption that each quasiparticle band is p-wave paired.
If the two types of spin-orbit couplings appear jointly, both the
singlet and triplet pairings are weakened and decreased down to
their minimum values in the equal-Rashba-Dresselhaus case.


\end{abstract}
\maketitle
\section{Introduction}
Superconductivity in materials without inversion symmetry has
attracted a lot of interests after the discovery of the heavy
fermion noncentrosymmetric (NCS) superconductor $\rm{CePt_{3}Si}$
\cite{bauer,frigeri,bin,bin2,linder,samo,mineev}. Due to the lack of
inversion symmetry, antisymmetric spin-orbit coupling (SOC) is
introduced \cite{frigeri,winkler}. There are two typical SOCs namely
the so-called Rashba \cite{rashba} and Dresselhaus
\cite{Dresselhaus} SOCs. The former is related to the microscopic
structural inversion asymmetry and can be described by the form
$H_{\mathrm{RSOC}}=\alpha (\sigma_y k_x-\sigma_x k_y)$ in a
two-dimensional (2D) system \cite{rashba,bychkov}, while the latter
arises due to the bulk inversion asymmetry in crystalline structures
and the interface inversion asymmetry, with the linear form
$H_{\mathrm{DSOC}}=\beta (\sigma_x k_x-\sigma_y k_y)$ in a 2D case
\cite{Dresselhaus,vardanyan,ganichev2003,ganichev2004}. $\alpha$,
$\beta$ are the coupling constants of Rashba and Dresselhaus terms,
respectively. SOC is crucial for the novel properties in NCS
superconductors \cite{frigeri}.

In most previous studies, people focus on the effect of Rashba type
SOC upon superconductivity
\cite{frigeri,bin,bin2,linder,samo,mineev,xu}. The Rashba SOC is
reported to induce spin splitting and mixing of the spin-singlet and
triplet pairings in a 2D superconducting system \cite{gorkov2001}.
Both the spin-singlet and triplet pairings are found to be enhanced
by Rashba SOC \cite{linder}. In addition to the Rashba type SOC,
Dresselhaus SOC also contributes to the band splitting.
Consequently, the similar effect on superconductivity is expected in
the presence of Dresselhaus SOC. However, the details in this case
are still unknown. Moreover, the combination of Rashba and
Dresselhaus SOCs has been realized in semiconductor quantum wells
\cite{ganichev2004,koralek2009} and ultra-cold atoms
\cite{lin2011,galitski}. It is found that a lot of interesting
physical phenomena appear in the presence of both Rashba and
Dresselhaus couplings \cite{chaoli,dell'anna2012,zhouli}. While in
the NCS superconductors, the combined effect of Rashba and
Dresselhaus SOCs on superconductivity remains open.

In this paper, we introduce a simple model to describe a 2D NCS
superconducting system in the presence of both Rashba and
Dresselhaus SOCs. Then the combined effect of the two SOCs on
superconductivity can be investigated in this model. The pairing
order parameters are solved self-consistently within the mean-field
theory. It is shown that an admixture of spin-singlet and triplet
pairing can be induced by either pure Rashba/Dresselhaus SOC or the
combination of the two SOCs. When Rashba and Dresselhaus SOCs are
present solely, both the spin-singlet and triplet pairings are
enhanced by increasing SOC. While in the case of Rashba and
Dresselhaus SOCs are mixed, the two pairing gaps are weakened
continuously with increasing Dresselhaus component and reduced to
their minimum values in equal-Rashba-Dresselhaus case
$(\alpha=\beta$).

\section{The model}
We start from the normal state Hamiltonian in the presence of both
Rashba and Dresselhaus couplings as follows
\begin{eqnarray}
\label{eq.1}
H_{N}=\sum_{{\bf k},s}\varepsilon_{{\bf k}}c_{{\bf k}s}^\dagger c_{{\bf k} s}+H_{soc},
\end{eqnarray}
with
\begin{eqnarray}
H_{soc}&=\sum_{{\bf k},ss'}\left\{\alpha (\sigma_y k_x-\sigma_x k_y) \right.\nonumber \\
&\left.+\beta (\sigma_x k_x-\sigma_y k_y)\right\}_{ss'}
c_{{\bf k}s}^\dagger c_{{\bf k} s'},
\label{eq.2}
\end{eqnarray}
where $\varepsilon_{{\bf k}}=\frac{{\bf k}^2}{2m}-\mu$ is the
spin-independent single electron kinetic energy measured relative to
the chemical potential $\mu$. $c_{{\bf k}s}^\dagger (c_{{\bf k}
s'})$ is the creation (annihilation) operator of electron and
$s,s'=\uparrow,\downarrow$ are spin indices. $\alpha$ and $\beta$
are the Rashba and Dresselhaus SOC strength parameters,
respectively. ${\bf k}=(k_x,k_y)$ is the 2D electron wave vector,
and $\sigma_{x}$, $\sigma_{y}$ are the Pauli matrices.

By introducing an angle $\theta$ which denotes the strength ratio
between Rashba and Dresselhaus SOCs \cite{vardanyan,yangzh2008},
the Eq.~(\ref{eq.2}) can be rewritten as
\begin{eqnarray}
\label{eq.3}
H_{soc}=\sum_{{\bf k},ss'}\gamma \left(\tilde{\sigma}_y k_x-\tilde{\sigma}_x k_y \right)_{ss'}
c_{{\bf k}s}^\dagger c_{{\bf k} s'},
\end{eqnarray}
with $\gamma=\sqrt{\alpha^2+\beta^2}$, and $\alpha=\gamma \cos{\theta}$,
$\beta=\gamma \sin{\theta}$,
$
\tilde{\sigma}_x=\left(  \begin{array}{ccc} 0 & e^{-i\theta} \\ e^{i\theta} & 0 \end{array} \right)
$,
and $\tilde{\sigma}_y=\left(  \begin{array}{ccc} 0 & -ie^{i\theta} \\ ie^{-i\theta} & 0  \end{array} \right)$.
Then the Hamiltonian in Eq.~(\ref{eq.1}) reads
\begin{equation}
\label{eq.4}
H_{N}=\sum_{{\bf k},s}\varepsilon_{{\bf k}}c_{{\bf k}s}^\dagger c_{{\bf k} s}+\sum_{{\bf k},ss'}\gamma \left(\tilde{\sigma}_y k_x-\tilde{\sigma}_x k_y \right)_{ss'}
c_{{\bf k}s}^\dagger c_{{\bf k} s'}.
\end{equation}

Applying the diagonalization procedure with Bogliubov transformation
in Eq.~(\ref{eq.4}), we arrive at
\begin{eqnarray}
\label{eq.5}
H_{N}=\sum_{{\bf k}\lambda} \xi_{\lambda}({\bf k}) a_{{\bf k}\lambda}^\dagger a_{{\bf k}\lambda},
\end{eqnarray}
where $a_{{\bf k}\lambda}^\dagger (a_{{\bf k} \lambda})$ is the
creation (annihilation) operator of quasiparticle, and $\lambda=\pm$
labels the SOC lifted quasiparticle band. $\xi_{\lambda }({\bf
k})=\varepsilon_{{\bf k}}-\lambda \gamma |{\bf k}|
\varsigma(\theta,\phi_{\bf k})$ is the energy dispersion in each
quasiparticle band with $\varsigma(\theta,\phi_{\bf
k})=\sqrt{1-\sin{2\theta}\sin{2\phi_{\bf k}}}$ and $|{\bf
k}|=\sqrt{k_x^2+k_y^2}$. Also we get the following unitary
transformations
\begin{equation}
\label{eq.6}
\begin{array}{ccc}
C_{{\bf k}\uparrow}&=&\frac{1}{\sqrt{2}} a_{{\bf k}+}+\frac{1}{\sqrt{2}} e^{i \eta({\bf k},\theta)} a_{{\bf k}-}, \\
C_{{\bf k}\downarrow}&=&\frac{1}{\sqrt{2}} a_{{\bf k}-}-\frac{1}{\sqrt{2}} e^{-i \eta({\bf k},\theta)} a_{{\bf k}+},
\end{array}
\end{equation}
with $e^{i \lambda\eta({\bf k},\theta)}=\frac{i\lambda e^{i\lambda \theta}\sin{\phi_{{\bf k}}}
+e^{-i\lambda \theta}\cos{\phi_{{\bf k}}}}{\varsigma(\theta,\phi_{\bf k})}$, and $\tan{\phi_{{\bf k}}}=k_{x}/k_{y}$.

\begin{figure}[ht]
  \centering
  \includegraphics[width=0.45\textwidth]{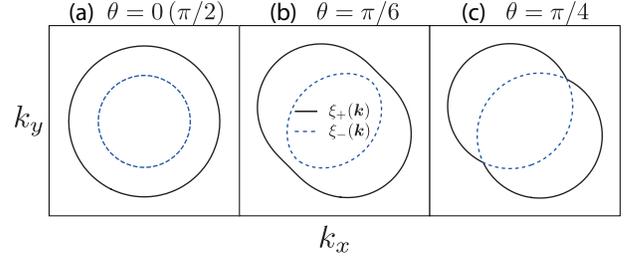}
  \vspace{-3mm}
  \caption{(Color online) The Fermi contour in the presence of Rashba
  and Dresselhaus SOCs with different $\theta$.
  (a) $\theta=0(\pi/2)$ represents only Rashba or Dresselhaus SOC.
  (b) $\theta=\pi/6$ stands for the combination of Rashba and Dresselhaus
  ones and $\alpha>\beta$.
  (c) $\theta=\pi/4$ shows the equal-Rashba-Dresselhaus case, $\alpha=\beta$.}
  \vspace{-3mm}
  \label{Fig.1} %% label for entire figure
\end{figure}

At zero temperature, the two quasiparticle bands are filled up to
the same Fermi energy level $\epsilon_{F}$, but with different Fermi
wave vectors. There are two Fermi contour lines corresponding to two
different dispersions $\xi_{\lambda}({\bf k})$ as shown in
Fig.~\ref{Fig.1}. For the system displaying pure Rashba ($\theta=0$)
or pure Dresselhaus ($\theta=\pi/2$) SOC, the Fermi contour lines
show similar isotropic concentric circles [see Fig.~\ref{Fig.1}(a)].
Rashba and Dresselhaus terms are found to play different roles on
the spin orientations in ${\bf k}$-space
\cite{ganichev2003,ganichev2004}, however it is not considered in
this paper. In the presence of both Rashba and Dresselhaus
couplings, the two Fermi contour lines are anisotropic and
non-equivalent along $[1 1 0]$ and $[\bar{1} 1 0]$ directions as
plotted in Figs.~\ref{Fig.1}(b) and~\ref{Fig.1}(c). Especially, when
$\alpha=\beta$, the two Fermi contour lines touch at $[1 1 0]$
direction, displayed in Fig.~\ref{Fig.1}(c). It is shown that
spin-splitting vanishes along a certain direction in this case
\cite{catalina}, and nontrivial physical properties can be expected.

In the strong SOC limit \cite{samo,mineev}, $k_{B}T_{C}\ll\gamma<\mu
$, the theory of NCS superconductor is analogous to that of
ferromagnetic superconductors \cite{jian}, and then only intra-band
pairing is allowed to occur in the same quasiparticle band
\cite{linder,samo,mineev}. Consequently, a p-wave pairing is
considered to be occur in each quasiparticle band, and the pairing
Hamiltonian can be written as
\begin{eqnarray}
\label{eq.7}
H_{sc}=\frac{1}{2N}\sum_{{\bf k},{\bf k'},\lambda}V_{\lambda}({\bf k},{\bf k'})
a_{{\bf k}\lambda}^\dagger a_{-{\bf k}\lambda}^\dagger a_{-{\bf k'}\lambda} a_{{\bf k'}\lambda},
\end{eqnarray}
where $N$ is the number of ${\bf k}$ points. We set the pairing
potential $V_{\lambda}({\bf k},{\bf k'})=-V_{\lambda}(-{\bf k},{\bf
k'})=-V_{\lambda}({\bf k},-{\bf k'}) =-V e^{i\lambda\left( \eta({\bf
k},\theta)- \eta({\bf k'},\theta)\right)}$ as analyzed in previous
studies \cite{linder,linder2}, and $\Delta_{\lambda}({\bf
k})=-\lambda \Delta_{\lambda} e^{i\lambda \eta({\bf k},\theta)}$. In
the weak coupling approach, the pairing interaction is nonzero only
inside the thin shells of width $\omega_{c}$ in the vicinity of
Fermi surface, and $\omega_{c}$ is chosen to be same in each
quasiparticle band.

To obtain the pairing order parameters, we define Green's functions
$G_{\lambda}({\bf k},\tau-\tau')=-\langle T_{\tau} a_{\lambda}({\bf
k},\tau) a_{\lambda}^\dagger({\bf k},\tau')\rangle$,
$F_{\lambda}({\bf k},\tau-\tau')=\langle T_{\tau} a_{\lambda}({\bf
k},\tau) a_{\lambda}(-{\bf k},\tau')\rangle$. The motion equations
of Green's functions in each band can be written as follows
\begin{equation}
\label{eq.8}
\begin{array}{lll}
\{i\omega_n-\xi_{\lambda}({\bf k})\}G_{\lambda}({\bf k},\omega_n)+\Delta_{\lambda}({\bf k})F_{\lambda}^\dagger(-{\bf k},\omega_n)=1,\\
\{i\omega_n+\xi_{\lambda}({\bf k})\}F_{\lambda}^\dagger(-{\bf k},\omega_n)+\Delta_{\lambda}^\dagger({\bf k}) G_{\lambda}({\bf k},\omega_n)=0.
\end{array}
\end{equation}
Then the obtained Green's functions read
\begin{eqnarray}
\label{eq.9}
G_{\lambda}({\bf k},\omega_n)=\frac{i\omega_n+\xi_{\lambda}({\bf k})}{(i\omega_n)^2-E_{\lambda}^2({\bf k})}, \\
F_{\lambda}^\dagger({\bf k},\omega_n)=\frac{-\Delta_\lambda^*({\bf k})}{(i\omega_n)^2-E_{\lambda}^2({\bf k})},
\end{eqnarray}
where $E_{\lambda}({\bf k})=\sqrt{\xi_{\lambda}^2({\bf
k})+\Delta_{\lambda}^2({\bf k})}$ is the quasiparticle excitation
energy for each band. The pairing order parameter in each
quasiparticle band is defined as
\begin{eqnarray}
\label{eq.10}
\Delta_{\lambda}({\bf k})=-\frac{1}{N}\sum_{{\bf k'}}V_{\lambda}({\bf k}{\bf k'})F_{\lambda}({\bf k'},0).
\end{eqnarray}
The chemical potential $\mu$ is determined from the particle number
density
\begin{eqnarray}
\label{eq.11}
n=\frac{1}{N}\sum_{{\bf k}}(\langle n_{{\bf k}\uparrow}\rangle+\langle n_{{\bf k}\downarrow}\rangle).
\end{eqnarray}

With $\langle n_{k\lambda}\rangle
=\frac{1}{2}-\frac{\xi_{\lambda}({\bf k})}{2E_{\lambda}({\bf
k})}\tanh{\frac{E_{\lambda}({\bf k})}{2 k_{B}T}}$, the order
parameters equations should satisfy
\begin{eqnarray}
\label{eq.12}
\Delta_\lambda({\bf k})=-\frac{1}{N}\sum_{{\bf k'}}V_{\lambda}({\bf k}{\bf k'})
\frac{\Delta_\lambda({\bf k'})\tanh{\frac{
E_{\lambda}({\bf k'})}{2k_{B}T}}}{2E_{\lambda}({\bf k'})},
\end{eqnarray}
\begin{eqnarray}
\label{eq.13}
n=\frac{1}{N}\sum_{{\bf k}}\left\{1-\frac{\xi_{+}({\bf k})}{2E_{+}({\bf k})}
\tanh{\frac{E_{+}({\bf k})}{2k_{B}T}} \right. \nonumber \\
\left.-\frac{\xi_{-}({\bf k})}{2E_{-}({\bf k})}
\tanh{\frac{ E_{-}({\bf k})}{2k_{B}T}}\right\}.
\end{eqnarray}

In the 2D case, the summations over ${\bf k}$ space in
Eqs.~(\ref{eq.12}) and (\ref{eq.13}) can be converted into continuum
integrals over energy by $\sum_{{\bf k}}=\frac{N}{(2\pi)^2}\int d^2
{\bf k}=\frac{N}{(2\pi)^2}\int k dk d\varphi$, where $k$ is the
magnitude of the momentum ${\bf k}$ and $\varphi$ is the polar
angle. The unit of the energy can be scaled by the factor
$\frac{\hbar^2 (2\pi n)}{2m}$ and the particle number density is set
as $n=1$ for half-filling. Accordingly, Eqs.~(\ref{eq.12}) and
(\ref{eq.13}) can be rewritten in the zero temperature limit with
$\tanh{\frac{E_{\lambda}({\bf k})}{2k_{B}T}}\rightarrow 1$ as
following form

\begin{eqnarray}
\label{eq.14}
\overline{\Delta}_{\lambda}=\frac{\overline{V}}{8\pi}\int_{\overline{\epsilon}_{\lambda}-\overline{\omega}_{c}}
^{\overline{\epsilon}_{\lambda}+\overline{\omega}_{c}}d\overline{\epsilon}\int_{0}^{2\pi}d\varphi
\frac{ \overline{\Delta}_{\lambda}}{\overline{E}_{\lambda}},
\end{eqnarray}

\begin{eqnarray}
\label{eq.15}
1=\frac{1}{8\pi}\int_{0}^{\infty}d\overline{\epsilon}\int_{0}^{2\pi}d\varphi
\left(2-\frac{\overline{\xi}_{+}}{\overline{E}_{+}}
-\frac{\overline{\xi}_{-}}{\overline{E}_{-}}\right),
\end{eqnarray}
where
$\overline{E}_{\lambda}=\sqrt{\overline{\xi}_{\lambda}^2+\overline{\Delta}_{\lambda}^2}$,
and
$\overline{\xi}_{\lambda}=\overline{\epsilon}-\overline{\mu}-\lambda
\overline{\gamma}_{\mathrm{soc}} \varsigma(\theta,\varphi)
\sqrt{\overline{\epsilon}}$ with
$\varsigma(\theta,\varphi)=\sqrt{1-\sin{2\theta}\sin{2\varphi}}$,
and $\varphi=\frac{\pi}{2}-\phi_{\bf k}$. The combined SOCs strength
is scaled as
$\overline{\gamma}_{\mathrm{soc}}=\sqrt{2\overline{\gamma}^2 m}$,
and $\overline{\epsilon}_{\lambda}=\frac{1}{2}(
\overline{\gamma}^2_{\mathrm{soc}} \varsigma^2(\theta,\varphi)+ 2
\overline{\mu}+\lambda \sqrt{4 \overline{\gamma}^2_{\mathrm{soc}}
\mu \varsigma^2(\theta,\varphi)+
\overline{\gamma}^4_{\mathrm{soc}}\varsigma^4(\theta,\varphi)}$. The
dimensionless pairing potential $\overline{V}$ is defined as
$\overline{V}=V*(\frac{\hbar^2 (2\pi n)}{2m})^{-1}$, and the
dimensionless energies $\overline{\epsilon}$, $\overline{\mu}$,
$\overline{\Delta}_{\lambda}$ and
$\overline{\gamma}^2_{\mathrm{soc}}$ are defined analogously. The
energy cutoff is chosen as $\overline{\omega}_{c}=0.01$.

When transformed into electron space using the unitary
transformations Eq.~(\ref{eq.6}), the pairing Hamiltonian
Eq.~(\ref{eq.7}) is an admixture of s-wave and p-wave pairing terms,
\begin{eqnarray}
H_{sc}&=&\frac{1}{4N}\sum_{{\bf k},{\bf k'},\sigma}V_{\sigma}({\bf k},{\bf k'})
c_{{\bf k}\sigma}^{\dagger}c_{-{\bf k}\sigma}^{\dagger}c_{-{\bf k}'\sigma}c_{{\bf k}'\sigma} \nonumber \\
&-&\frac{1}{2N}\sum_{{\bf k},{\bf k'},\sigma}V c_{{\bf k}\sigma}^{\dagger}c_{-{\bf k},-\sigma}^{\dagger}c_{-{\bf k}',-\sigma}c_{{\bf k}',\sigma},
\end{eqnarray}
where $V_{\sigma}({\bf k},{\bf k'})=-V e^{i\sigma\left( \eta({\bf
k}, \theta)- \eta({\bf k'}, \theta) \right)}$ is the electron p-wave
pairing potential and $\sigma=\pm1$ represents the electron spin.
The superconducting order parameters in electron space are also an
admixture of spin-singlet and spin-triplet ones accordingly, and can
be expressed as \cite{linder}
\begin{equation}
\label{eq.16}
\begin{array}{ccc}
\Delta_{\uparrow\uparrow}({\bf k})&=&-\frac{e^{i\eta({\bf k},\theta)}}{2}\left(\overline{\Delta}_{+}+\overline{\Delta}_{-}\right),\\
\Delta_{\downarrow\downarrow}({\bf k})&=&\frac{e^{-i\eta({\bf k},\theta)}}{2}\left(\overline{\Delta}_{+}+\overline{\Delta}_{-}\right),\\
\Delta_{\uparrow\downarrow}&=&\frac{1}{2}\left(\overline{\Delta}_{+}-\overline{\Delta}_{-}\right).
\end{array}
\end{equation}
From the equations above, we can calculate the effect of Rashba and
Dresselhaus SOCs on electron superconducting order parameters in
both spin-singlet and triplet channels.

\section{Results and Discussions}

\begin{figure}[ht]
  \centering
  \includegraphics[width=0.45\textwidth,keepaspectratio=true]{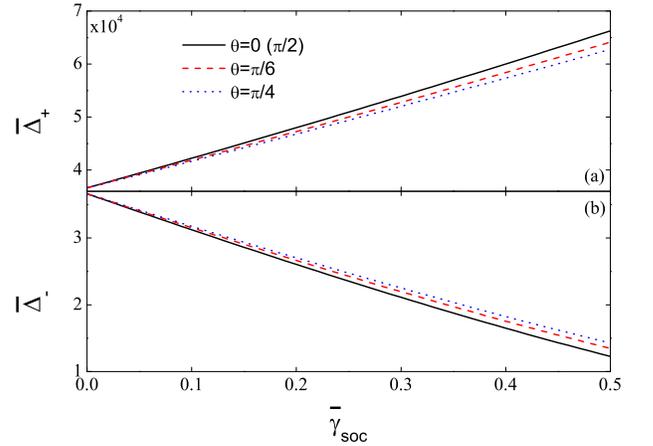}
  \vspace{-3mm}
  \caption{(Color online) The dependencies of the quasiparticle superconducting order parameters
  $\overline{\Delta}_{+}$ (a), $\overline{\Delta}_{-}$ (b) on SOC strength $\overline{\gamma}_{\mathrm{soc}}$ are plotted
  at $\overline{T}=0$ and $\overline{V}=0.5$.
  The angle $\theta=0,\ \pi/6, \ \pi/4$ represent $\beta=0$,
  $\alpha>\beta$ and $\alpha=\beta$, respectively.}
  \vspace{-3mm}
  \label{Fig.2} %% label for entire figure
\end{figure}

Firstly, we can calculate the quasiparticle pairing gaps
$\overline{\Delta}_{+}$ and $\overline{\Delta}_{-}$
self-consistently by solving Eqs.~(\ref{eq.14}) and (\ref{eq.15}).
As shown in Fig.~\ref{Fig.2}(a), no matter Rashba and Dresselhaus
SOCs are present solely ($\theta=0,(\pi/2)$) or jointly
($\theta=\pi/6,\pi/4$), $\overline{\Delta}_{+}$ is always enhanced
by SOC, while $\overline{\Delta}_{-}$ is weakened as SOC increased
displaying in Fig.~\ref{Fig.2}(b). The role of SOC in quasiparticle
pairings analogous to that of the ferromagnetism in ferromagnetic
superconductors \cite{jian}. When Rashba and Dresselhaus SOCs are
mixed ($\theta=\pi/6,\pi/4$), $\overline{\Delta}_{+}$ is reduced
with the increase of Dresselhaus component, while
$\overline{\Delta}_{-}$ is strengthened. The mixed SOC seems to
soften the effect of pure Rashba or Dresselhaus SOC on the
quasiparticle pairings.

With $\overline{\Delta}_{+}$ and $\overline{\Delta}_{-}$ in hand and
according to Eq.~(\ref{eq.16}), we can investigate the electron
pairings upon increasing SOC. From the Eq.~(\ref{eq.16}), it is
clearly shown that $|\Delta_{\uparrow\uparrow}({\bf
k})|=|\Delta_{\downarrow\downarrow}({\bf k})|$. For convenience, we
set
$\overline{\Delta}_{\uparrow\uparrow}=|\Delta_{\uparrow\uparrow}({\bf
k})|+|\Delta_{\downarrow\downarrow}({\bf k})|$,
$\overline{\Delta}_{\uparrow\downarrow}$ as the electron
spin-triplet and singlet pairing components, respectively. In the
absence of SOC, $\overline{\Delta}_{+}=\overline{\Delta}_{-}$, the
spin-singlet pairing component
$\overline{\Delta}_{\uparrow\downarrow}=0$, i.e., the
superconducting state is a pure triplet one.

\begin{figure}[ht]
  \centering
  \includegraphics[width=0.45\textwidth,keepaspectratio=true]{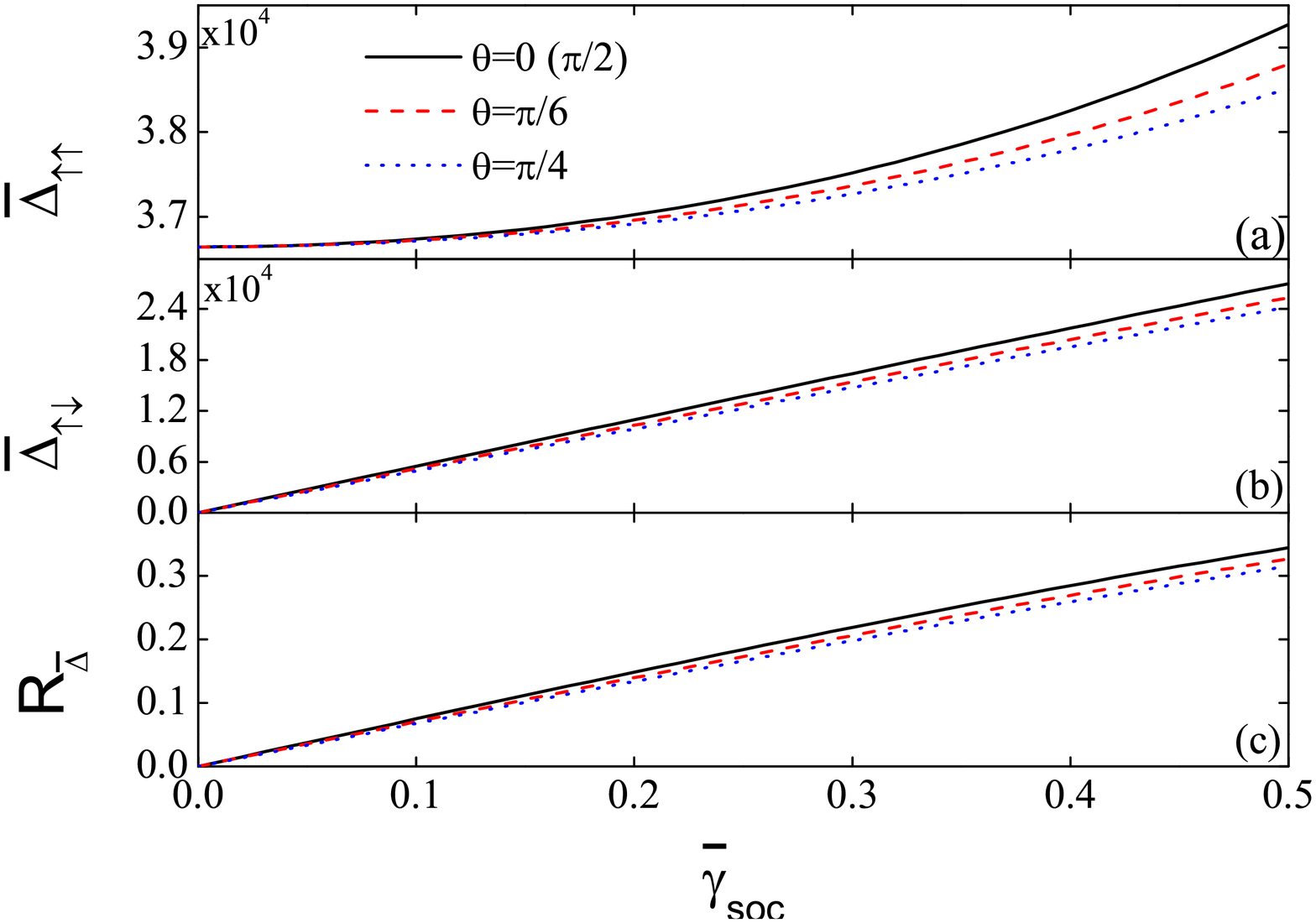}
  \vspace{-3mm}
  \caption{(Color online) Plots of the electron pairing order parameters
 spin-triplet component $\overline{\Delta}_{\uparrow\uparrow}$ (a),
  spin-singlet component $\overline{\Delta}_{\uparrow\downarrow}$ (b) and
  the ratio of spin-singlet to spin-triplet component $R_{\overline{\Delta}}$ (c)
  as functions of $\overline{\gamma}_{\mathrm{soc}}$ with different $\theta$.
  All the parameters are solved at $\overline{T}=0$ and $\overline{V}=0.5$.}
  \vspace{-3mm}
  \label{Fig.3} %% label for entire figure
\end{figure}

The order parameters $\overline{\Delta}_{\uparrow\uparrow}$,
$\overline{\Delta}_{\uparrow\downarrow}$ and the ratio
$R_{\overline{\Delta}}=\overline{\Delta}_{\uparrow\downarrow}/\overline{\Delta}_{\uparrow\uparrow}$
are plotted as a function of $\overline{\gamma}_{\mathrm{soc}}$ in
Fig.~\ref{Fig.3}. When pure Rashba or Dresselhaus SOC is present
($\theta=0, (\pi/2)$), both the spin-triplet and singlet pairing
components are enhanced with increasing SOC. The only difference
between Rashba and Dresselhaus SOCs on the pairing gaps is related
to a phase factor as shown in Eq.~(\ref{eq.16}). For the pure Rashba
SOC system, $e^{i\lambda\eta({\bf k},\theta)}=e^{i\lambda \phi_{\bf
k}}$, while in the pure Dresselhaus SOC case $e^{i\lambda\eta({\bf
k},\theta)}=-i\lambda e^{-i\lambda \phi_{\bf k}}$. However, in this
paper we only focus on the magnitude of the order parameter. In this
case, the pure Rashba and Dresselhaus SOCs are considered to result
in the similar effect on the pairing gaps. In the presence of both
Rashba and Dresselhaus SOCs ($\theta=\pi/6, \pi/4$ ), the combined
SOCs also enhance the spin-singlet and triplet pairings. Moreover,
for a fixed $\overline{\gamma}_{\mathrm{soc}}$, the electron
pairings in both spin-triplet and spin-singlet channels are weakened
by increasing Dresselhaus component (see Figs.~\ref{Fig.3}(a) and
\ref{Fig.3}(b)). As displayed in Fig.~\ref{Fig.3}(c), the ratio
$R_{\overline{\Delta}}$ increases from zero as a function of
$\overline{\gamma}_{\mathrm{soc}}$ in all the cases $\theta=0,
(\pi/2)$, $\theta=\pi/6$ and $\theta=\pi/4$, implying that
spin-singlet pairing is induced and increased by pure Rashba or
Dressehaus SOC and the mixed SOCs.

\begin{figure}[ht]
  \centering
  \includegraphics[width=0.45\textwidth,keepaspectratio=true]{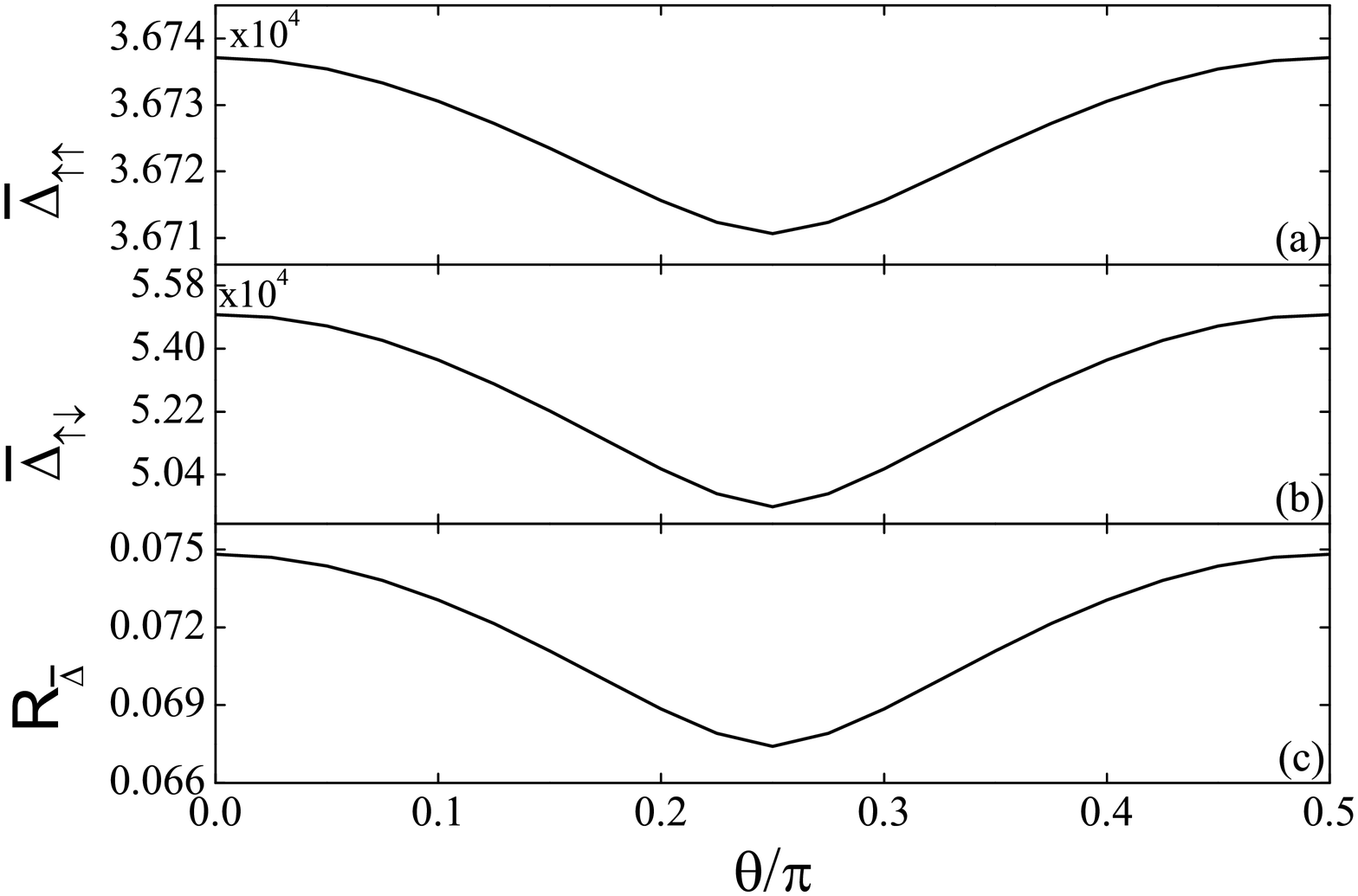}
  \vspace{-3mm}
  \caption{Shown are the plots of the
  electron pairing gaps $\overline{\Delta}_{\uparrow\downarrow}$ (a),
  $\overline{\Delta}_{\uparrow\uparrow}$ (b)
  and the ratio $R_{\overline{\Delta}}$ (c) as functions of $\theta$ at $\overline{T}=0$,
  $\overline{V}=0.5$ and $\overline{\gamma}_{\mathrm{soc}}=0.1$.}
  \vspace{-3mm}
  \label{Fig.4} %% label for entire figure
\end{figure}

Fig.~\ref{Fig.4} displays the pairing parameters
$\overline{\Delta}_{\uparrow\uparrow}$,
$\overline{\Delta}_{\uparrow\downarrow}$ and the ratio
$R_{\overline{\Delta}}$ as the variation of $\theta$. As $\theta$
increases, the Dresselhaus component in the combined SOCs is
increased. All the $\overline{\Delta}_{\uparrow\uparrow}$,
$\overline{\Delta}_{\uparrow\downarrow}$ and $R_{\overline{\Delta}}$
are shown to reduce continuously with increasing $\theta$, and
decrease down to their minimum values right at $\theta=\pi/4$
($\alpha=\beta$). In the case of $\alpha=\beta$, the splitting of
the Fermi contour lines vanishes at a certain direction as shown in
Fig.~\ref{Fig.1}(c). The minimum values may correspond to this
splitting vanishing. As $\theta$ increases further, the order
parameters and the ratio $R_{\overline{\Delta}}$ rise again. When
$\theta$ increases to $\theta=\pi/2$ ($\alpha=0$), only Dresselhaus
SOC establishes in the system, and the values of all the order
parameters are maximal and equal to those when $\theta=0$.

\section{Summary}
In summary, the combined effect of the Rashba and Dresselhaus SOCs
on superconductivity is investigated within the mean-field theory.
Either the Rashba or Dresselhaus SOC can mix the spin degree of
freedom and thus may give rise to two nondegenerate quasiparticle
bands. This explains why spin-singlet paring in the electron band
can be induced from the spin-triplet paring of quasiparticles. Both
spin-singlet and triplet electron-paring are enhanced with the
strength of SOC increased. However, if the two SOCs are combined but
the effective coupling strength keeps a constant, we find that both
spin-singlet and triplet electron-paring are weakened and decrease
down to their minimum values in the case that the Rashba and
Dresselhaus SOCs are equally mixed. In this case, the Fermi contour
lines of the two quasiparticle bands touch at a certain direction.

\section*{Acknowledgements}
This work is supported by the National Natural Science Foundation of
China (Grant No.~11274039), the Specialized Research Fund for the
Doctoral Program of Higher Education (No.~20100006110021) and the
Fundamental Research Funds for the Central Universities of China. QG
acknowledges helpful discussions with Prof. John Chalker and is
grateful for the support from the China Scholarship Council and the
hospitality of the Rudolf Peierls Centre for Theoretical Physics,
University of Oxford.

\end{document}